\documentstyle[preprint,aps]{revtex}

\begin{document}

\preprint{FIMAT-8/95}

\title{Three-dimensional effects on the electronic structure of
quasiperiodic systems}

\author{Enrique Maci\'{a}\thanks{Also at the
Instituto de Estudios Interdisciplinares, El Guijo, Z4 Galapagar,
E-28260 Madrid, Spain.} and Francisco Dom\'{\i}nguez-Adame}

\address{Departamento de F\'{\i}sica de Materiales,
Facultad de F\'{\i}sicas, Universidad Complutense,\\
E-28040 Madrid, Spain}

\maketitle

\begin{abstract}

We report on a theoreticl study of the electronic structure of
quasiperiodic, quasi-one-dimensional systems where fully three
dimensional interaction potentials are taken into account.  In our
approach, the actual physical potential acting upon the electrons is
replaced by a set of nonlocal separable potentials, leading to an
exactly solvable Schr\"odinger equation.  By choosing an appropriate
trial potential, we obtain a discrete set of algebraic equations that
can be mapped onto a general tight-binding-like equation.  We introduce
a Fibonacci sequence either in the strength of the on-site potentials or
in the nearest-neighbor distances, and we find numerically that these
systems present a highly fragmented, self-similar electronic spectrum,
which becomes singular continuous in the thermodynamical limit.  In this
way we extend the results obtained so far in one-dimensional models to
the three-dimensional case. As an example of the application of the
model we consider the chain polymer case.

\end{abstract}


\narrowtext

\section{Introduction}

The study of transport properties of self-similar aperiodic structures
has been a continuously growing research field during the last decade.
On the one side, the remarkable discovery of the quasicrystalline phase
\cite{Sh,Goldman} gave rise to a considerably amount of theoretical work
on transport properties of Fibonacci lattices, regarded as archetypical
models of quasiperiodic systems in one dimension (1D)
\cite{Suther,Wurtz,Goda,Severin,Capaz,Newman,PLA}.  On the other side,
since the first growth of semiconducting quasiperiodic superlattices
\cite{Merlin,Todd}, it has been progressively realized that electronic
systems arranged according to a substitution sequence should offer
interesting possibilities for technological applications.  In addition,
recent progresses in diffusion-controlled aggregation on surfaces,
allowing to growth 1D chains of atoms on a substrate \cite{Nature},
along with proposals of nanoscale devices based on atomic switching in
atomic wires \cite{Wada}, span considerably the interest of the {\em
quasiperiodic order} notion \cite{PRE} to properly describe transport
properties of quasiperiodic systems (QS) at the nanostructure level.

It is well known that, for 1D systems, quasiperiodicity leads to highly
fragmented energy spectra that are Cantor sets with zero Lebesgue
measure in the thermodynamical limit \cite{Luck}. This exotic spectrum
determines the existence of electron wave functions that are neither
extended in the Bloch sense nor localized, but display dramatic
fluctuations over the whole system.  As a consequence, transport
properties exhibit a peculiar behavior even when finite temperature
effects are taken into account \cite{PRBFIBO}. To the date, however,
most of the theoretical work on the transport properties of QS breaks
into two main kinds of 1D models, namely, discrete tight-binding models
and continuous Kronig-Penney models.  Albeit purely 1D models frequently
render as adequate approximations to more complex three-dimensional (3D)
systems, it is also true that low dimensional treatments could give rise
to spurious phenomena that are not met in going to higher dimensions.
Then the question as to whether results obtained for purely 1D systems
can be extended to more realistic, fully 3D models becomes very
appealing from both theoretical and technological points of view.  In
fact, if the purported fragmentation of the electronic spectrum of QS is
destroyed by 3D effects, then theoretical results obtained so far about
its related characteristic transport properties are of no practical
relevance and possible technological application of these interesting
systems fade away.

In this work we address the task of elucidating whether the remarkable
properties of 1D QS are inherent to the low dimensionality of the models
or, on the contrary, they should be expected in 3D systems as well.  To
this end, we introduce a completely general model to study electronic
properties in 3D QS. Our approach is based on the so-called nonlocal
(separable) potential (NLP) method, in which the actual potential at
each site of an arbitrary system is replaced by a projective operator
\cite{Knight,Sievert,MP}. As an interesting working example, we shall
consider in detail the case of trans-polyacetylene.  The different
nonlinear excitations present in polymers in general, like solitons,
polarons, and bipolarons are not included in the present study.  If one
is interested in the effect of these types of excitations, it would be
necessary to resort to many-body Hamiltonians like the
Su-Schrieffer-Heeger or the Peierls-Hubbard ones (see
Ref.~\onlinecite{a} and references therein).  On the other hand, what we
are interested in is on 3D electron dynamics in quasiperiodic systems,
and in showing them as much clear as possible.  Thus, it is reasonable
to focus only on linear excitations.  The fact that we use parameters
for trans-polyacetylene later is because we have obtained them in the
framework of our nonlocal potential model with great accuracy \cite{MP},
and therefore they are already available to present an example of the
orders of magnitude to be expected in polymer applications.  In this way
we show that characteristic features of 1D QS are still present when 3D
interaction potentials are considered.  The rest of the paper is
organized as follows.  In Sec.~II we present our model, summarize
previous work of us \cite{MP} that is necessary for a better
understanding of the present paper, and obtain exact expressions for the
physical magnitudes of interest in the study of general QS. Afterwards,
in Sec.~III we undertake a detailed study of the electronic properties
of 3D QS. We obtain numerical results demonstrating the existence of a
highly-fragmented, self-similar electronic spectrum, being singular
continuous in the thermodynamical limit.  Section IV concludes the paper
with a brief discussion on the physical relevance of the obtained
results.

\section{The NLP approach to the study of quasiperiodic systems}

The solution of the Schr\"odinger equation for multicentre interactions
is of interest in condensed matter physics as well as in atomic and
molecular physics.  As it is well known, such solution is expected to
involve enormous difficulties since, in most cases, prohibitively
cumbersome calculations are required.  During last years several methods
have been proposed to solve this problem.  Among them, the NLP approach
is the natural generalization of the Kronig-Penney model to the 3D case.
This method yields an exactly solvable Schr\"odinger equation from which
the electron energy can be readily obtained without tedious and
elaborated calculations.  What is more important, it is always possible
to find a NLP (or a sum of them) able to reproduce any set of given
electronic states \cite{Larry} and, consequently, there is no
theoretical limitation to the numerical accuracy with which physical
results can be obtained.  We shall open this section with a brief
account of the general aspects of the NLP approach and, subsequently, we
will apply it to the description of QS.

\subsection{Schr\"odinger equation for NLP}

The Schr\"odinger equation for NLP reads as follows \cite{MP} (we take
$\hbar=m=1$ hereafter)
\begin{equation}
\left( {\bf p}^2-2E \right) \psi ({\bf r}) =
\sum_k\> \lambda_k V(|{\bf r}-{\bf R}_k|) \int \> d^3r'
V(|{\bf r}'-{\bf R}_k|) \psi ({\bf r}' ),
\label{1}
\end{equation}
where ${\bf R}_k$ denotes the position of each atomic site and
$\lambda_k$ is the corresponding coupling constant.  It is usual to
assume, as a first approximation, that the potential function $V$ is
spherically symmetric, although other symmetries can be also considered
within the NLP approach. By Fourier transforming we have
\begin{equation}
\psi ({\bf p}) = \left( {1\over p^2-2E} \right)
\sum_k\> \lambda_k V(p)\exp (-i{\bf p}\cdot{\bf R}_k) \chi_k,
\label{2}
\end{equation}
where
\begin{equation}
\chi_k=\int \>d^3p\,V^{*}(p)\exp(i{\bf p}\cdot{\bf R}_k)\psi({\bf p}).
\label{3}
\end{equation}
Here $\psi({\bf p})$ and $V(p)$ denote the Fourier transforms of $\psi$
and $V$, respectively.  The coefficients $\chi_k$ are related to the
wave function in real space, and we will discuss their meaning once we
have specified the potential $V(p)$.  Inserting (\ref{2}) in
(\ref{3}) and performing the angular integration we obtain the following
set of algebraic equations
\begin{equation}
\chi_k= 4\pi\sum_j\> \lambda_j\, \int_0^\infty \>dp\,
\frac{p^2|V(p)|^2}{p^2-2E}\
\frac{\sin pR_{kj}}{pR_{kj}}\chi_j,
\label{5}
\end{equation}
where $R_{kj}\equiv |{\bf R}_k-{\bf R}_j|$ and it is understood that the
factor $(\sin pR_{kj})/pR_{kj}$ is replaced by $1$ when $k=j$.

\subsection{Application to quasi-one-dimensional lattices}

The set of algebraic equations given in (\ref{5}) provide a completely
general solution of the Schr\"odinger equation for multicentre
interactions in three dimensions, as soon as the shape of the potential
$V(p)$ is specified.  Hence, the crucial point within the NLP approach
is to choose an appropriate potential that reproduces the observed
energy values for the physical system being considered.  In this work we
will be concerned with quasi-one-dimensional systems where an array of
3D potentials are arranged along a straight line.  We have shown that
the electronic band structure of quasi-one-dimensional polymers, such as
trans-polyacetylene, can be quite accurately obtained making use of
this approach and taking surface $\delta$-function interactions of the
form \cite{MP}
\begin{equation}
V(r)={1\over r^2}\,\delta (r-R), \hspace{7mm}
V(p)=\sqrt{2\over \pi} \, {\sin pR\over pR}.
\label{6}
\end{equation}
Physically this potential describes a highly localized force field which
vanishes everywhere except on a spherical shell of radius $R$ around the
lattice site.  In this particular case Eq.~(\ref{5}) takes the form
\begin{equation}
\chi_k= {\lambda_k\over R^2}\, A(E)\chi_k + \sum_{j\neq k}\>
{\lambda_j\over R^2}\, B_{kj}(E)\chi_j,
\label{7}
\end{equation}
where for brevity we have defined
\begin{eqnarray}
A(E)&=&{2\pi\over\kappa}\left[1-\exp(-2\kappa R)\right],\nonumber \\
B_{kj}(E)&=&{2\pi \exp (-\kappa R_{kj}) \over \kappa^2R_{kj}}
\left( \cosh 2\kappa R-1\right).
\label{8}
\end{eqnarray}
The corresponding integrals have been performed for the case of
interest, namely $E<0$, and then $\kappa\equiv \sqrt{-2E}$ is a real
parameter.  Notice that the larger the distance between site $k$ and
$j$, the smaller the corresponding coefficient $B_{kj}(E)$.  In other
words, such coefficients are rapidly decreasing functions of $R_{kj}$
provided that $\kappa$ is not very small (deep potentials).  This is
usually a good approximation in most cases of practical interest and,
for these systems, it is reasonable to assume that only
nearest-neighbor interactions are significant.  Thus
\begin{equation}
\chi_k= {\lambda_k\over R^2}\, A(E)\chi_k +
{\lambda_{k+1}\over R^2}\, B_{k\,k+1}(E)\chi_{k+1} +
{\lambda_{k-1}\over R^2}\, B_{k\,k-1}(E)\chi_{k-1}.
\label{9}
\end{equation}
At this point it becomes convenient to provide some physical insight
into the coupling constants $\lambda_j$ and, to this end, we must
consider the single site case for a moment.  When only one
atom at ${\bf R}_k=0$ is present in the system, the NLP Schr\"odinger
equation reduces to
\begin{equation}
\left( {\bf p}^2-2E \right) \psi ({\bf r}) =
\lambda_k V(r) \int \> d^3r'\,V(r')\psi({\bf r}'),
\label{1bis}
\end{equation}
from which we get
\begin{equation}
\frac{1}{4\pi\lambda_k} = \int_0^\infty \>dp\,
\frac{p^2|V(p)|^2}{p^2-2E_k}.
\label{5bis}
\end{equation}
This relation enables us to link the coupling constant of any arbitrary
site of the system with its single bound state energy level
$E_k=-\kappa_k^2/2$, an experimentally measurable quantity.  Making use
of expressions (\ref{6}) and (\ref{5bis}) we get $\lambda_k
A(E_k)/R^2=1$.  For short-ranged potentials we can expand $A(E_k)$ in
powers of $R$ obtaining
\begin{equation}
\lambda_k \sim {R\over 4\pi}(1+\kappa_k R).
\label{10}
\end{equation}
In this way we can estimate the most appropriate values of the coupling
constants from experimental data.  Now we go back to our general
treatment.  Inserting (\ref{10}) in (\ref{9}) and taking the
limit $R\to 0$ in such a way that $E_k$ remains constant, we obtain the
following tight-binding equation for the coefficients $\chi_k$
\begin{equation}
(\kappa-\kappa_k)\chi_k=
\frac{\exp (-\kappa R_{k\,k+1})}{R_{k\,k+1}}\chi_{k+1}+
\frac{\exp (-\kappa R_{k\,k-1})}{R_{k\,k-1}}\chi_{k-1}.
\label{11}
\end{equation}
Note that the corresponding transfer integrals decrease exponentially on
the distance between nearest-neighbors, as it should be expected, and
their functional form is typical of s-orbitals.  In addition they depend
explicitly on the electron energy.  Therefore, at this stage of our
analysis, the mathematical treatment of the system remains fully 3D
although the system is quasi-one-dimensional.  Before proceeding, we
wish to clarify the physical meaning of $\chi_k$.  From their definition
in Eq.~(\ref{3}) and the Parseval identity we have
\begin{equation}
\chi_k=\int \>d^3r\,V(r)\psi({\bf r}+{\bf R}_k),
\label{13}
\end{equation}
with $V(r)=r^{-2}\delta (r-R)$. In the limit $R\to 0$ one gets
$V(r)\to r^{-2}\delta (r)=\delta({\bf r})$. Therefore, in this
limiting case
\begin{equation}
\chi_k=\psi({\bf R}_k).
\label{14}
\end{equation}
Thus we see that $\chi_k$ is nothing but the value of the electron
wave function at site ${\bf R}_k$, which is of course the quantity of
interest.

\subsection{The NLP approach for quasiperiodic quasi-one-dimensional
systems}

We now proceed introducing quasiperiodicity in our lattice model.  This
can be accomplished in two different ways.  On the one hand, we can
assume the nearest-neighbor distance to be constant and consider two
kind of different basic units, say A and B, arranged quasiperiodically
along the chain.  Examples of these systems may be provided by binary
Fibonacci quasicrystals or multilayered heterostructures.  On the other
hand, we can assume all the sites to be the same and vary the distance
between them in a quasiperiodic fashion.  In this case, the sites can be
occupied by either single atoms (self-assembled aggregates), molecules
(homopolymers) or an assembly of crystalline monolayers (superlattices).
When applied to the tight-binding Eq.~(\ref{11}), the first option
corresponds to on-site models, whereas the second one describe transfer
models.

During the last years several quasiperiodic arrangements, based on the
application of substitution rules, have been considered in the
literature.  In this work we will focus on the Fibonacci arrangement as
a canonical example of QS which has been experimentally probed in a
variety of situations \cite{Laruelle,Nakayama,Kono,Tuet} in the
knowledge that our treatment can be similarly applied to other
self-similar aperiodic realizations in a straightforward manner.  In
general, a Fibonacci chain of order $l$ is generated from two basic
units $A$ and $B$ by successive applications of the substitution $A\to
AB$ and $B\to A$ yielding a sequence of the form $ABAABABA \ldots$ This
sequence comprises $F_{l-1}$ elements $A$ and $F_{l-2}$ elements $B$,
where $F_l$ is the {\em l\/}th Fibonacci number given by the recurrent
law $F_l=F_{l-1} + F_{l-2}$ with $F_0=F_1=1$.  As $l$ increases the
ratio $F_{l-1}/F_l$ converges toward $\tau= (\sqrt{5}-1)/2$, which is
known as the inverse golden mean.

The electron dynamics of this system can be conveniently studied by
means of the well-known transfer-matrix techniques. To this end, we cast
Eq.~(\ref{11}) into the matrix form
\begin{equation}
\left( \begin{array}{c} \chi_{k+1} \\ \chi_{k} \end{array} \right) =
\left( \begin{array}{cc} \alpha_k & -\beta_k \\ 1 & 0 \end{array}
\right)
\left( \begin{array}{c} \chi_{k} \\ \chi_{k-1} \end{array} \right)
\equiv P_{k}
\left( \begin{array}{c} \chi_{k} \\ \chi_{k-1} \end{array} \right),
\label{21}
\end{equation}
where $\alpha_k=(\kappa- \kappa_k)R_{k\,k+1} \exp (\kappa R_{k\,k+1})$
and $\beta_k=(R_{k\,k+1}/R_{k\,k-1}) \exp (\kappa R_{k\,k+1}-
\kappa R_{k\,k-1})$. The transfer matrix of the whole system is then
found as \begin{equation}
T(N)=\prod_{k=N}^1\>P_k,
\label{22}
\end{equation}
which relates the wave function at both edges of the system.  Taking
into account the fact that $T(N)=P_N\, T(N-1)$ with $T(0)$ the $2 \times
2$ unity matrix, we
find the following recurrence relations to compute the transfer-matrix
elements
\begin{eqnarray}
T_{11}(n) &=& \alpha_n T_{11}(n-1)- \beta_n T_{11}(n-2),  \nonumber
\\ T_{12}(n) &=& \alpha_n T_{12}(n-1)- \beta_n T_{12}(n-2),
\nonumber \\ T_{21}(n) &=& T_{11}(n-1),  \nonumber  \\
T_{22}(n) &=& T_{12}(n-1), \hspace{1 true cm} n=2,3\, \cdots N.
\label{25}
\end{eqnarray}
with the initial conditions $T_{ij}(0)=\delta_{ij}$, $T_{11}(1)=
\alpha_1$, $T_{12}(1)=-\beta_1$, $T_{21}(1)=1$ and $T_{22}(1)=0$.  Here
$N$ is a Fibonacci number indicating the number of basic units of the
system.  Some physically relevant magnitudes can be readily obtained
from the transfer-matrix $T(N)$.  Although the information that anyone
of these magnitudes can provide isolately is not conclusive, when
grouped together they produce a clear picture on the nature of the
electronic states \cite{prbito}.  The Lyapunov coefficient, representing
the growth rate of the wave function, is nothing but the inverse of the
localization length.  As a consequence, delocalization of electronic
wave function is seen through the decrease of this parameter.  It can be
computed as \cite{Borland}
\begin{equation}
\Gamma (E) = \left( {1\over N} \right)
\left( T_{11}^2(N)+T_{12}^2(N) +T_{21}^2(N)+ T_{22}^2(N) \right).
\label{26}
\end{equation}
The fragmentation scheme and the number of states in each subband is
characterized by means of the integrated density of states (IDOS) per
unit length.  Finally, considering periodic boundary conditions at both
edges of the system, the following condition for an energy to be in an
allowed band is obtained
\begin{equation}
\left({1\over 2}\right)\left|\mbox{Tr}\left[ T(N) \right]\right|\leq 1.
\label{27}
\end{equation}
Making use of this condition the equivalent bandwidth (defined below) of
the spectrum can be readily determined and, from its scaling properties
with the system size, we can infer the spectral signature of the
electronic spectrum.  All these expressions are very simple and suitable
for an efficient numerical treatment.  We will now evaluate them for
several interesting cases.

\section{Results and discussions}

For convenience, we shall describe the results obtained for transfer and
on-site models separately.

\subsection{Quasiperiodic transfer systems}

As an illustrative example we shall consider a quasi-one-dimensional
array of CH units roughly representing the trans-polyacetylene chain.
There are two parameters that can be varied in a transfer model, namely,
the values of the short, $R_S$, and long, $R_L$, bond lengths between
neighboring units.  Two different sets of short and long alternating
bonds have been considered describing the extreme cases of weak
($R_L=1.43\,$\AA, $R_S=1.36\,$\AA) and strong ($R_L=1.54\,$\AA ,
$R_S=1.34\,$\AA) polymer stretching \cite{Grant}.  Making use of
previous results \cite{MP} we can estimate the range of on-site energies
able to reproduce the experimentally observed band edges in
trans-polyacetylene depending on its stretching state.  In this way we
obtain $E_0 \sim -6.0$\, eV to be an appropriate on-site energy on the
average.  Chains containing up to N=$F_{15}=987$ CH units have been
studied numerically making use of the expressions derived in the
previous section.  In Fig.\ 1 we show the fragmentation of the
$\pi$-band as a function of the Fibonacci number obtained from condition
(\ref{27}) in the case of strong stretching.  Only short approximants of
the Fibonacci sequence are displayed since, on increasing N, the
spectrum becomes so fragmented that it becomes very difficult to observe
minor features in the plot.  We have carefully analysed, however,
spectra up to $N=987$ in order to determine the spectrum splitting
pattern.  In this way we have confirmed that the single band
corresponding to the periodic case splits into four main subbands which
further split into smaller subsubbands.  As soon as a quasiperiodic
arrangement of the conjugated bonds is introduced, the position and
widths of the main subbands of the spectrum converge rapidly to stable
values on increasing the polymer length.  This behavior implies that
the global structure of the electronic spectrum can be obtained, in
practice, by considering very short approximants to the infinite,
strictly quasiperiodic realization.  In fact, this is the case for
system size as short as $N=F_{8}=34$ in the considered model.  This
characteristic has been referred to as the {\em asymptotic stability} of
the spectrum \cite{PRBFIBO}.

{}From a statistical point of view, valuable information can be extracted
from the way in which the electronic spectrum splits on increasing the
number of units in the chain.  Our previous experience in the study of
1D systems indicates that the IDOS per unit length is a powerful tool to
study the occupation of the different subbands of the fragmented
spectrum.  In Fig.\ 2 we present a typical IDOS evaluated by means of
node counting.  Its stair-case shape provides additional support on the
self-similar splitting scheme just reported on.  In fact, the overall
tetrafurcation pattern of the spectrum is characterized by the presence
of four main subbands, labeled $a$, $b$, $c$, and $d$ in Fig.\ 2,
separated by well-defined plateaus.  Inside each main subband the
fragmentation proceeds obeying a trifurcation scheme which can be
quantitatively described as follows.  The number of subsubbands in each
main subband is given by $N_a=F_{l-2}$, $N_b=F_{l-3}$, $N_c=F_{l-4}$,
and $N_d=F_{l-3}$ in agreement with distribution rules reported for 1D
tight-binding models \cite{PRBFIBO,Riklund}.  We define the {\em
occupation} of a main subband as the ratio between the number of states
in that subband and the total number of electronic states, this is to
say, $q_i=N_i/N$ ($i=a,b,c,d$).  Since each site in the carbon backbone
contributes with one electronic state to the energy spectrum, this
occupation can be directly measured from the heights of the
characteristic steps appearing in the IDOS.  In this way we have
determined that $q_a=\tau^2$, $q_b=\tau^3$, $q_c=\tau^4$, and
$q_d=\tau^3$, within an error less than $0.1 \%$, for a wide collection
of system realizations. it is interesting to note that these occupations
correspond to a nonequilibrium level distribution for the system at zero
temperature, in agreement with previous results for 1D QS \cite{PRE}.

\subsection{Quasiperiodic on-site systems}

There are two parameters that can be varied in an on-site model, namely,
the bound state levels of the composing units $E_A$ and $E_B$.  For the
sake of illustration we shall consider a quasi-one-dimensional array of
CH units describing {\em chemically doped} trans-polyacetylene.  By this
we mean that the native polymer has been subjected to a chemical process
consisting of either a charge transfer reaction by oxidizing or reducing
agents or an acidic treatment in which protons attach to certain sites
of the chain.  In our model unchanged polymer units will be located in
sites $A$ whereas units doped in this way are placed quasiperiodically
at sites $B$.  Assuming an uniform carbon-carbon bond length of
$L=1.39\,$\AA \cite{Grant}. we obtain that $E_A\sim -6.0\,$eV reproduces
quite well the observed band edges in trans-polyacetylene \cite{MP}.  On
the other side, the value of $E_B$, which describes the effect of doping
on the different units, will be varied in the range from $-6.0\,$eV up
to $-12.0\,$eV in order to ascertain its influence on the electronic
structure of the system.

Making use of the value $R_{k\,k\pm 1}=L$ and defining $\rho\equiv
\kappa L$ and $\rho_k\equiv \kappa_k L$ for the sake of brevity, we can
express the tight-binding equation (\ref{11}) in the form
\begin{equation}
(\rho-\rho_k)e^\rho\chi_k=\chi_{k\,k+1}+\chi_{k\,k-1}.
\label{12}
\end{equation}
We have checked that only one band is placed in the negative-energy
interval in the case of native trans-polyacetylene with an uniform
carbon-carbon bond length and that, when doping is uniformly introduced,
this band becomes deeper and narrower on decreasing the value of the
on-site energy associated to the doped sites.  The general behavior of
the $\pi$-band as a function of this on-site energy is shown in Fig.\ 3.
Deviation from native polymer state can be quantitatively measured by
means of the ratio $\alpha=E_B/E_A$, hereafter called {\em doping
parameter}. The more distant is $\alpha$ from unity, the more important
is the effect of doping.

Using of condition (\ref{27}), we have found that, as soon as
the native polymer is chemically modified at sites arranged
quasiperiodically, the $\pi$-band splits into four main subbands which
further split into smaller subsubbands.  The observed splitting
hierarchy is completely analogous to that shown in Fig.~1.  A
characteristic feature of Fibonacci systems, which has been thoroughly
purported in the study of 1D systems, concerns the self-similar pattern
exhibited by their energy spectra.  In Fig.\ 4 we show a typical plot of
the Lyapunov coefficient as a function of the electron energy.  In this
plot the maxima indicate the position of the minigaps determining the
spectrum fragmentation discussed before.  By inspecting this plot, it is
clear that the whole spectrum corresponding to a short Fibonacci
approximant is mapped onto a small portion of the spectrum.

It has been rigorously shown that, in the thermodynamical limit, 1D
Fibonacci systems present a singular continuous electronic spectrum
\cite{Ghez}.  In order to estimate the spectral type associated to our
system, we have computed the so-called equivalent bandwidth $S$, defined
as the sum of all allowed subbands.  As can be expected from the
Cantor-like nature of Fibonaccian spectra, $S$ vanishes as the polymer
size increases according to a power law of the form
$S=F_l^{-\beta(\alpha)}$.  In Fig.\ 5 we show the dependence of the
equivalent bandwidth on the polymer size for different values of the
doping parameter $\alpha$.  We have confirmed that $S$ obeys a power law
behavior even when high doping parameters are considered (up to
$\alpha=2$).  According to earlier results \cite{Kohmoto}, such a
power-law scaling is characteristic of a singular continuous spectrum
for which all the wave functions are critical, i.\ e., regarding
localization properties the functions are neither exponentially
localized nor extended in the Bloch sense.  Hence we are led to the
conclusion that 3D effects do not change the critical nature of the
electronic states in QS.  To get more information on how the electronic
spectra behaves in approaching the thermodynamical limit we undertook
the study of the dependence of the critical exponent $\beta$ on the
ratio $\alpha=E_B/E_A$ with a fixed value $E_A=-6.0\,$eV.  Results shown
in Fig.\ 6 evidence that $\beta$ raises linearly with $\alpha$, hence
suggesting that the electronic spectrum is fragmented more rapidly on
increasing the chemical diversity of the chain.  This is to be expected
since the fragmentation of the $\pi$-band arises as a consequence of
loss of quantum coherence, and this effect is enhanced as the on-site
energies ($E_A$ and $E_B$) become more and more different due to the
doping process.

\section{conclusions}

In this paper we have considered electron dynamics in
quasi-one-dimensional Fibonacci systems.  Our procedure, based on the
NLP approach, allows us to carry out a fully three dimensional analysis
of the considered system.  It is important to realize that this
technique can be applied to any desired level of precision.  In
addition, the exact solution can be found for an arbitrary NLP, as we
actually demonstrated [see Eq.~(\ref{5})] by means of the Fourier
transform which, in turn, is completely equivalent to the use of the
Green function formalism.  We have considered surface $\delta$-function
interactions with vanishing radius since this potential gives very
accurate results in the study of trans-polyacetylene \cite{MP}. We have
focused on the analysis of quasiperiodic arrangements where atoms are
placed in a straight line.  Nevertheless, the model can be easily
extended to other possible geometric configurations of interest as, for
example, zig-zag chains whose $sp^2$ hybridization geometry is
explicitly considered, or branched atomic wires as those obtained for
diffusion-controlled aggregates.

In this work we demonstrate that all those features reported as to being
typical of quasiperiodic (Fibonacci) systems in one dimension can be
extended to the three dimensional case as well.  In this way we provide
substantial support to the interest of these novel systems from an
applied viewpoint.  In particular we wish stress the fact that the most
characteristic property of electronic energy spectrum of QS, namely, its
highly fragmented self-similar structure, is expected to be observed
also in more realistic (3D) realizations.  This result strongly suggests
that the peculiar nature of the spectra associated to quasiperiodic
systems has nothing to do with the number of spatial dimensions
considered in the interaction potential but it is directly related to
the topological order implied by the way in which the basic units are
arranged in the system.  A further elaboration of this point, dealing
with three dimensional quasiperiodic Schr\"odinger Hamiltonian operators
from a mathematical point of view will be, therefore, very appealing.

Albeit we have focused on a particular realization, trans-polyacetylene,
attending to the practical interest of this system, our main results are
very general in character and, in fact, encompass a broad class of
quasiperiodically ordered systems, as we have discussed previously.  As
a consequence, we can confidently drawn the following conclusions.  In
the first place, it is worth noticing that the self-similar
fragmentation of the spectrum is a {\em robust} property of 3D QS.
Indeed, we have confirmed it in two different scenarios, namely, extreme
cases of stretched bonding (transfer models) and wide range of chemical
doping (on-site models).  In both cases the hierarchy of the splitting
spectrum pattern remains unchanged and the gap labelling ordering
preserved.  In the second place, we observe that the asymptotic
stability of the electronic spectrum, which was previously reported in
the study of quasiperiodic superlattices \cite{PRBFIBO}. can be extended
to a wide class of QS. This result has important consequences from a
practical viewpoint as it indicates that typically fractal features can
be observed in the transport properties of relatively short QS, which
are within the domain of actual technological capabilities.  In the
third place, we have shown that the equivalent bandwidth of the
electronic spectrum can be modified in a precise way by properly
choosing the sample size and the chemical diversity, hence allowing for
a certain kind of engineering of the related transport properties.
Finally, we confirm that the systems which we are dealing with deserve a
proper position in the orderings of matter by their own right.  In fact,
in a recent work \cite{PRE} we have reported that 1D QS are able to
encode more information, in the Shannon sense, than other usual
orderings of matter are able to do.  This feature is intimately related
to the fact that QS describe far from thermodynamical equilibrium
systems.  In this work we provide substantial support to this view,
since we are considering a polymer which is not in the equilibrium state
described by an alternating pattern of the conjugated bonds.  Instead,
we are dealing with a polymer with either a quasiperiodic arrangement of
short and long bonds or a quasiperiodic sequence of chemically doped
sites.  In both cases we are describing systems which are not in its
state of minimum energy.  Therefore, it is not surprising that the
corresponding IDOS display reversed level distribution at zero
temperature (see Fig.\ 2), a typical feature of nonequilibrium systems.
To the best of our knowledge the experimental realization of such a kind
of polymers has not been reported in the literature.  In this sense, we
expect the theoretical results we report in this work could receive
further attention by experimental researchers.

\acknowledgments

This work is supported by CICYT through project MAT95-0325.

\begin{figure}
\caption{Allowed subbands as a function of the Fibonacci order $l$ for a
QS polymer with $R_L=1.54\,$\AA, $R_S=1.34\,$\AA\ and $E=-6.0\,$eV}
\label{fig1}
\end{figure}

\begin{figure}
\caption{IDOS versus energy for a QS polymer with
$L_A=1.54\,$\AA, $L_B=1.34\,$\AA, $E=-6.0\,$eV and $N=F_{15}=987$.  The
main subbands are labelled by {\em a,b,c} and {\em d} along with their
respective occupations.  See the text for further details.}
\label{fig2}
\end{figure}

\begin{figure}
\caption{Allowed energies (shaded region) as a function of the on-site
energy in the native trans-polyacetylene.  The carbon-carbon bond length
is $L=1.39\,$\AA.}
\label{fig3}
\end{figure}

\begin{figure}
\caption{(a) Lyapunov coefficient as a function of energy for a
QS polymer with $L=1.39\,$\AA, $E_A=-6.0\,$eV and $E_B=-7.0\,$eV.
(b) An enlarged view of one of the main subbands illustrating its
self-similar nature.}
\label{fig4}
\end{figure}

\begin{figure}
\caption{Equivalent bandwidth $S$ as a function of the system size $N$
for $E_A=-6.0\,$eV and different values of $E_B$, indicated on each
curve.  The carbon-carbon bond length is $L=1.39\,$\AA.}
\label{fig5}
\end{figure}

\begin{figure}
\caption{Critical exponent $\beta$ as a function of the doping ratio
$\alpha=E_B/E_A$ with $E_A=-6.0\,$eV.  The carbon-carbon bond length is
$L=1.39\,$\AA.}
\label{fig6}
\end{figure}

\end{document}